%% file: main.tex
\documentclass[conference]{IEEEtran}
\IEEEoverridecommandlockouts
\usepackage{cite}
\usepackage{amsmath,amssymb}
\usepackage{graphicx}
\usepackage{xcolor}
\usepackage{colortbl}
\usepackage{tikz}
\usetikzlibrary{arrows.meta,decorations.pathmorphing,decorations.pathreplacing,positioning}
\usepackage{booktabs}
\usepackage{multirow}
\usepackage{url}
\usepackage[hidelinks]{hyperref}
\newcommand{\Pfail}{P_{\mathrm{fail}}}

\begin{document}

\title{Sense Less, Miss Less:\\ Evidence-Driven Sensing Schedules for ISAC}

\author{
\IEEEauthorblockN{Seyed Bagher Hashemi Natanzi\thanks{Corresponding author: snatanzi@wpi.edu}, Bo Tang}
\IEEEauthorblockA{Department of Electrical and Computer Engineering, Worcester Polytechnic Institute (WPI), USA}
\IEEEauthorblockA{Email: \{snatanzi, btang1\}@wpi.edu}
}

\maketitle
\maketitle
    \begin{tikzpicture}[remember picture, overlay]
        \node[anchor=north, yshift=-0.5cm] at (current page.north) {\fbox{\parbox{\textwidth}{\centering\small {\color{red}This work has been submitted to the IEEE Conference for possible publication. Copyright may be transferred without notice, after which this version may no longer be accessible.}}}};
    \end{tikzpicture}
\begin{abstract}
Integrated sensing and communication (ISAC) lets a base station detect human motion with the same signals it uses for data, but every sensing transmission takes resources from users. Sensing rarely while nothing moves saves resources, yet a brief hand gesture by an otherwise still person can then start and end between two transmissions and go unnoticed. We propose an evidence-driven schedule, based on data-efficient cumulative-sum (CuSum) change detection, that pauses sensing while measurements indicate no motion and senses continuously once they suggest motion. It is adapted to orthogonal frequency-division multiplexing (OFDM) sensing, where requests can be blocked by data traffic and measurements arrive late, and evaluated with an articulated human model of limb motion. In simulations without wall reflections, with the same monitoring budget and below one false alarm per minute, it detects 1.5--4.2 times as many limb movements within 200~ms as periodic sensing. A tuned two-rate trigger performs similarly, and the gain over periodic sensing persists, without retuning or cost matching, in a simulated room with wall reflections. Under a small budget, measurements should decide when to sense.
\end{abstract}


\section{Introduction}

Cellular networks are beginning to sense as well as communicate. In integrated sensing and communication (ISAC), a base station reuses the echoes of its own orthogonal frequency-division multiplexing (OFDM) transmissions as radar returns~\cite{liu2022isac}, and these echoes reveal not only where people are but how they move: each moving body segment adds its own Doppler shift, and the resulting time-varying pattern, the micro-Doppler signature, supports gesture and activity recognition~\cite{chen2006microdoppler,pegoraro2022sparcs}. A base station could thus notice a raised hand or a reach towards a machine without cameras or wearables.

Sensing, however, is not free. Every sensing transmission occupies radio resources, scheduler attention and processing that could otherwise serve users. The natural economy is to sense often when something happens and rarely when the scene is quiet, and recent work does exactly that: it adapts sensing resources to user motion~\cite{elase}, triggers sounding on demand alongside communication reference signals~\cite{jiang2026carts}, or chooses sampling patterns that preserve micro-Doppler quality under a budget~\cite{mason2025sampling}. These schemes target tracking accuracy, channel quality or spectrogram reconstruction.

Sensing rarely has a price that none of them measures. Consider a person standing still in front of a base station that has lowered its sensing rate accordingly. The person raises a hand; the movement lasts a fraction of a second and the torso does not move. If observations are sparse, the gesture can begin and end between two of them, or be recognised only after it is over, which is useless for an interface or a safety reaction. This paper therefore asks: given a limited monitoring budget, i.e., a limited number of sensing transmissions per second spent watching for events, \emph{when} should the base station look so that brief limb movements are still detected within a deadline?

Two obstacles stand in the way. The first is the target model. In the ISAC channel model of the 3rd Generation Partnership Project (3GPP), TR~38.901~\cite[Sec.~7.9.2]{3gpp38901}, a human is a single scattering point moving as a rigid body; a waving person with a still torso produces no limb Doppler at all (Fig.~\ref{fig:models}a), so studies built on it cannot even see the problem.

The second is the detector. Quickest change detection raises an alarm as soon as possible after the statistics of a measured signal change, subject to a limit on false alarms, and the cumulative-sum (CuSum) test~\cite{page1954cusum} is its classical tool. Data-efficient CuSum (DE-CuSum)~\cite{banerjee2013decusum} also decides when to observe, skipping observations while the evidence favours the unchanged state; related work treats transient changes~\cite{chandar2015transient}, the choice among sensing actions~\cite{veeravalli2024controlled}, and the trade-off between communication rate and detection delay in ISAC codeword design~\cite{seo2025qcd}. These results assume idealised observations. An OFDM base station sees the scene differently: Doppler needs several transmissions spread over tens of milliseconds, requests collide with data traffic, measurements arrive late, and breathing, posture and clutter produce evidence that resembles motion.

This paper closes both gaps by letting the evidence decide when to look. Throughout, a \emph{sensing occasion} is one sensing transmission, and a \emph{look} is a short group of occasions from which Doppler content can be estimated. The contributions are:
\begin{enumerate}
\item A formulation of transient limb-motion detection with deadline, monitoring-budget and false-alarm constraints under intermittent, delayed OFDM observations.
\item An articulated 12-segment human model that captures limb motion missing from the single-point target, with its channel cross-checked against the radar cross section (RCS) solver of the Sionna~RT ray tracer~\cite{hoydis2023sionnart}.
\item An adaptation of DE-CuSum to OFDM looks, with guidelines for their duration and spacing.
\item An evaluation fixed before testing (pre-registered) at matched monitoring cost and a common false-alarm target, extended by a two-rate trigger baseline, a sensing-rate sweep and a multipath transfer test.
\end{enumerate}
The results show that the timing of sensing matters as much as its amount: at matched monitoring cost, evidence-driven sensing detects 1.5--4.2 times as many brief limb movements in time as periodic sensing. When sensing is scarce, a base station should not look on a timetable; it should look when the evidence asks it to.\footnote{ \url{https://github.com/CLIS-WPI/isac-transient-motion-detection}}

\section{System Model}

\subsection{Monostatic OFDM Sensing}

We consider a full-duplex base station (gNB or gNodeB) performing monostatic OFDM sensing, i.e., transmitting and receiving at the same node, at carrier frequency $f_c$ with slot duration $T_s$. A sensing occasion consists of one OFDM symbol carrying a sensing reference signal on $K$ subcarriers distributed uniformly across bandwidth $B$. Self-interference is assumed to be suppressed. In slot $n$, the echo on subcarrier $k$ is
\begin{equation}
y_k[n] = g[n]\sum_{p} \alpha_p[n]\, e^{-j2\pi (f_c+f_k)\tau_p[n]} + w_k[n],
\label{eq:channel}
\end{equation}
where $f_k$ is the subcarrier frequency offset from $f_c$, $c_0$ is the speed of light, and $\lambda=c_0/f_c$. The sum runs over human and static clutter scattering points $p$, with round-trip delay $\tau_p=2R_p/c_0$ at range $R_p$. The radar-equation amplitude is $\alpha_p=\sqrt{\sigma_p}\,\lambda/((4\pi)^{3/2}R_p^2)$ for RCS $\sigma_p$. A slowly varying common gain $g[n]$ models residual background variation, and $w_k[n]$ is complex Gaussian noise.

\subsection{An Articulated Human}

\begin{figure}[t]
\centering
\includegraphics[width=\columnwidth]{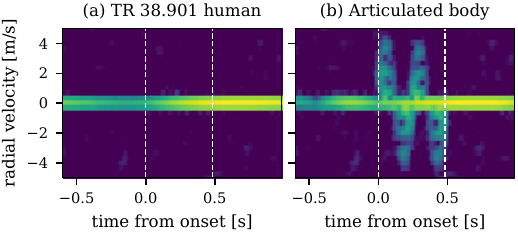}
\caption{Spectrogram of the same arm swing (torso still, peak limb speed 3.4~m/s) seen through (a) the single-point TR~38.901 human and (b) the articulated body. Dashed lines mark the start and end of the movement; colour shows echo power over a 40~dB range.}
\label{fig:models}
\end{figure}

The human is represented by $M$ prolate ellipsoidal segments with articulated motion of their centres and orientations, following established micro-Doppler simulation practice~\cite{boulic1990walking,chen2006microdoppler}. Each segment is represented by a scattering point at its centre with monostatic RCS
\begin{equation}
\sigma = \frac{\pi a^4 c^2}{\left(a^2\sin^2\theta + c^2\cos^2\theta\right)^2},
\label{eq:rcs}
\end{equation}
where $a$ is the radius, $c$ the half-length and $\theta$ the angle between the segment axis and the line of sight. The RCS therefore changes as a limb rotates. A \emph{target event} is a fast limb movement while the torso remains in place. Non-event motion includes breathing, postural sway, small joint jitter and nuisance movements, i.e., slow voluntary motions that are not target events, such as a head turn.

\subsection{Causal Observation Model}

Time is discretised into slots. The \emph{sensing policy}, the rule that decides when to request sensing occasions and when to raise an alarm, runs at the distributed unit (DU). Each request reaches the scheduler after a control delay and must target a slot at least a minimum lead time later. Requests targeting slots occupied by data traffic are rejected. Accepted measurements are delivered to the policy after a processing delay. The policy receives only delivered measurements and scheduler acceptances or rejections; it has no access to the underlying channel, noise realisations or event labels.

\subsection{Problem Statement}

Let an event begin at $t_0$ and end at $t_1$. An \emph{alarm} is the policy's declaration that a target event has started. Detection is successful if the first alarm occurs in $[t_0,\min(t_0+D,t_1)]$, where $D$ is the detection deadline. An event is missed if no qualifying alarm occurs within this interval, including when the alarm arrives after the movement ends. The failure probability $\Pfail$ is estimated by the fraction of missed events. Alarms outside $[t_0,t_1+G]$ are classified as false alarms, where $G$ is a post-event grace interval, and $r_{\mathrm{FA}}$ denotes the number of false alarms per minute of event-free time. Monitoring cost $C$ is the number of sensing occasions actually transmitted per second while watching for events. We seek to minimise detection failure subject to false-alarm and monitoring-cost constraints:
\begin{equation}
\min \ \Pfail \quad \text{s.t.} \quad r_{\mathrm{FA}} \le r_{\max}, \qquad C \le C_{\max},
\label{eq:problem}
\end{equation}
where $r_{\max}$ is the false-alarm limit. Each alarm starts a refractory period of duration $T_{\mathrm{ref}}$, during which monitoring pauses and no further alarm is raised; it also triggers $T_{\mathrm{svc}}$ of dense application-service sensing for the downstream application (e.g., gesture recognition). Monitoring cost $C$ excludes these service occasions; service cost is reported separately and included in the total-cost comparison in Section~V.

\section{Detection with Adaptive Observation}

\begin{figure}[t]
\centering
\resizebox{\columnwidth}{!}{\input{figures/fig_system.tex}}
\caption{Sensing configuration and adaptive observation. (a) A full-duplex gNB senses limb motion while the torso remains stationary. (b) Sensing occasions form looks; requests targeting data slots are rejected, and the adaptive schedule skips looks while its detection statistic $W$ (Section~\ref{sec:schedules}) is negative. (c) Processing and decision loop at the DU, including request and measurement delivery delays; LLR denotes the log-likelihood ratio of Section~\ref{sec:du}.}
\label{fig:system}
\end{figure}
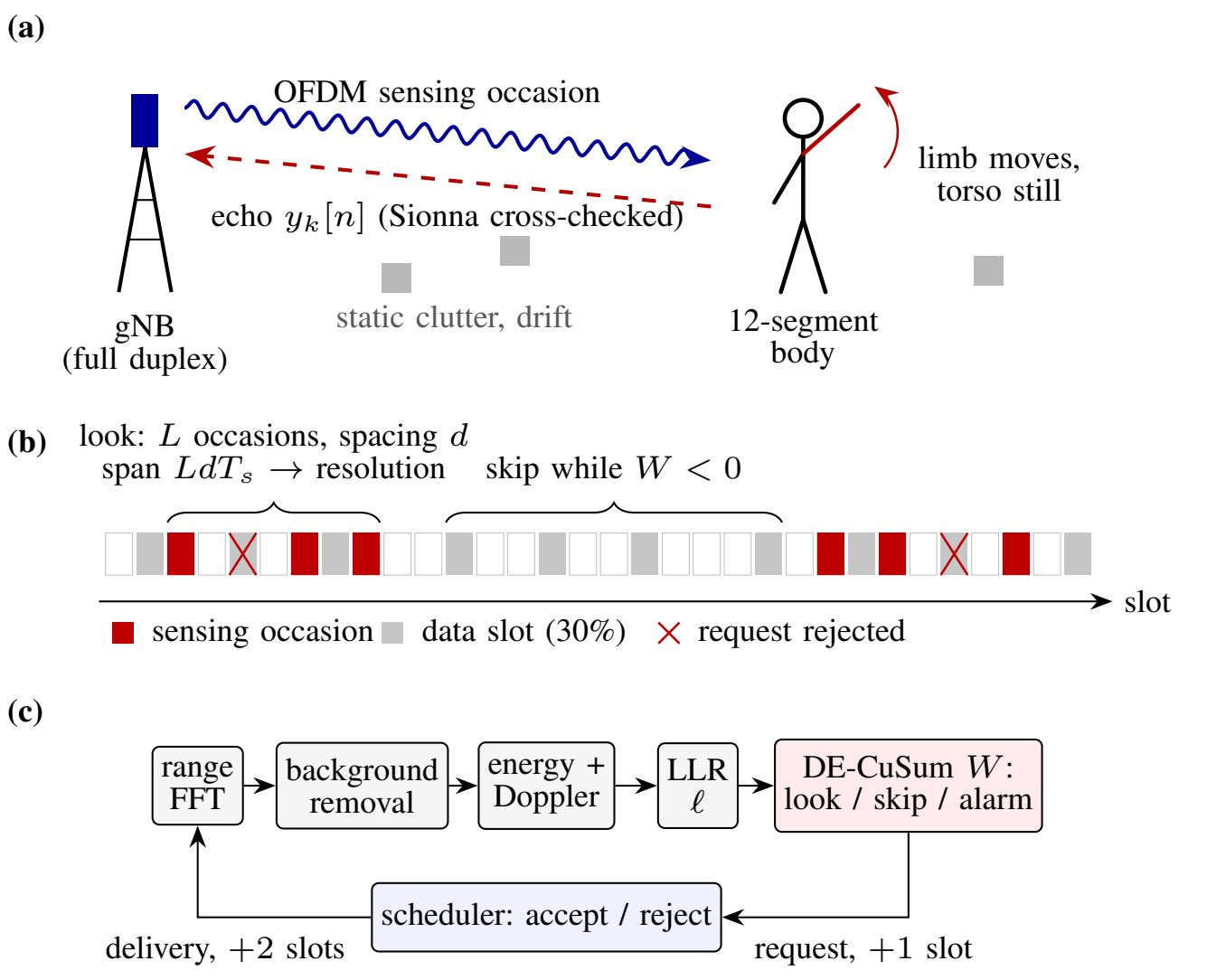

Fig.~\ref{fig:system} shows the sensing configuration and DU decision loop.

\subsection{Processing at the DU}\label{sec:du}

Each delivered measurement is transformed to range using an inverse fast Fourier transform (FFT) over the $K$ subcarriers. A background estimate is maintained for each range bin using exponential averaging with a 1~s time constant, adjusted for irregular sample spacing, and subtracted from the measurement.

For a window of $N_r$ residual samples $e_i$ acquired at times $t_i$, the DU computes two features. The first is residual energy after complex linear detrending, normalised by the noise power. Detrending suppresses breathing, sway and drift that are approximately linear over the observation window. The second is the fraction of spectral energy at $|f|>f_{\mathrm{cut}}$, using $S(f)=\big|\sum_i w_i e_i e^{-j2\pi f t_i}\big|^2$, where $w_i$ is a raised-cosine taper over the window. The spectrum is evaluated up to $1/(2\overline{\Delta t})$, with $\overline{\Delta t}$ denoting median sample spacing; this upper bound is a heuristic for irregular sampling, rather than an exact unambiguous Doppler band. The cut-off is $f_{\mathrm{cut}}=\max(15~\mathrm{Hz},\,1.5/T_{\mathrm{win}})$, capped at 80\% of that bound, where $T_{\mathrm{win}}$ is the window span, so that slow motion inside the main lobe at zero Doppler is not counted. Windows with fewer than four accepted samples produce no update.

Features come from the bin with the strongest residual. Class-conditional Gaussian models fitted on training data yield a window log-likelihood ratio (LLR) $\ell$, positive when the window favours an ongoing event and clipped to $[-8,8]$. All schedules share this pipeline.

\subsection{Looks and Their Design}\label{sec:looks}

Each look contains $L$ requested occasions spaced $d$ slots apart. Its nominal span $LdT_s$, i.e., the first-to-last interval $(L-1)dT_s$ plus one spacing, determines Doppler resolution, while spacing $dT_s$ determines the nominal unambiguous Doppler band $\pm1/(2dT_s)$ for the regularly spaced requests. Both duration and spacing must therefore be selected.

\subsection{Sensing Schedules}\label{sec:schedules}

Let $b$ denote the nominal fraction of slots requested for monitoring. The \emph{uniform} schedule requests one occasion every $1/b$ slots and computes $\ell$ over a sliding window of the last $N_w$ samples. The \emph{periodic burst} requests one look every $L/b$ slots and computes $\ell$ once per look. Both accumulate evidence in a detection statistic $W$ with the CuSum recursion~\cite{page1954cusum}, $W\leftarrow\max(0,W+\ell)$, and raise an alarm when $W$ exceeds the threshold $A$.

The \emph{two-rate trigger} requests one look every $P_{\mathrm{slow}}$ slots. If a look returns $\ell>\tau$, it requests $N_{\mathrm{hold}}$ consecutive looks before returning to the slow rate. It uses the same CuSum alarm rule.

The \emph{adaptive} schedule uses an OFDM adaptation of DE-CuSum~\cite{banerjee2013decusum}. Time is divided into epochs of one nominal look span, with update rule
\begin{equation}
W \leftarrow
\begin{cases}
\max(W+\ell,\,-h), & W\ge 0 \ \ \text{(take a look)},\\
\min(W+\mu,\,0), & W<0 \ \ \text{(skip)}.
\end{cases}
\label{eq:decusum}
\end{equation}
An alarm occurs when $W>A$. Negative evidence makes $W<0$ and suspends observation; the truncation level $h$ bounds how negative $W$ can become, and the recovery step $\mu$ sets how fast $W$ returns to zero during skipped epochs. Once $W$ climbs back to zero, looks resume and continue while $W\ge0$. Realised monitoring cost depends on these parameters, the evidence and accepted requests.

Five modifications account for the OFDM observation process. First, each observation is a look containing multiple sensing occasions, with duration and spacing selected as described in Section~\ref{sec:looks}. Second, $W$ is updated only after all accepted measurements in the look are delivered, so request and processing delays contribute to alarm latency. This window-level update is analogous to the subblock CuSum of~\cite{seo2025qcd}, which updates its statistic once per block of symbols. Third, requests targeting data slots are rejected, and $\ell$ is computed from the accepted samples using a model fitted for that look design; if fewer than four occasions of a look are accepted, $W$ keeps its value and a new look is requested in the next epoch.

Fourth, background averaging accounts for the actual sample spacing, and post-alarm service measurements refresh the estimate after skipped observations. Fifth, the alarm procedure includes a refractory period; post-alarm service cost is accounted for separately from monitoring cost. No optimality guarantee is claimed for this adaptation.

\section{Simulation Setup and Results}

\subsection{Simulation Setup}\label{sec:setup}

The base station is 2.5~m high, with $f_c=3.5$~GHz, 30~kHz subcarrier spacing and $T_s=0.5$~ms. Each occasion uses $K=32$ subcarriers uniformly across $B=100$~MHz (nominal range resolution 1.5~m). Six static clutter points lie at 1--10~m. Gain $g[n]$ has amplitude and phase deviations of $2\times10^{-3}$; noise gives a 5~dB per-subcarrier signal-to-noise ratio (SNR) for a broadside torso at 5~m.

The $M=12$ segments represent the torso, head, upper arms, forearms, hands, thighs and shins. Each \emph{episode} is a simulated 5~s scene with one person; episodes sample height 1.55--1.95~m, range 3--7~m and facing angle within $\pm60^\circ$. Arm swings comprise 70\% of events (one or two cycles); others are leg movements. Events last 0.3--0.8~s with peak limb speeds 1.5--4.5~m/s. Background motion includes breathing (2--6~mm), sway, small joint jitter and an average of 1.2 nuisance motions per episode, such as slow arm adjustment, weight shift or head turn. Motion-script labels are independent of detector outputs.

The policy runs every four slots (2~ms). A request issued in slot $n$ reaches the scheduler in slot $n{+}1$, may target slot $n{+}2$ or later, and its acceptance or rejection is visible to the policy from slot $n{+}2$. Data traffic blocks 30\% of slots; requests to these slots are rejected. Measurements arrive two slots after acquisition. Noise and slot availability are fixed by episode, slot and subcarrier, so different schedules observing the same slot receive identical measurements, which makes comparisons paired.

The detection deadline is $D=200$~ms, the post-event grace interval is $G=0.5$~s, and the false-alarm limit is $r_{\max}=1~\mathrm{min}^{-1}$. Each alarm starts a $T_{\mathrm{ref}}=1$~s refractory period, whose first $T_{\mathrm{svc}}=200$~ms carry dense service sensing. Refractory time counts as event-free time for $r_{\mathrm{FA}}$; the event and grace interval do not.

For cross-checking, each segment and clutter point was a Sionna~RT~2.2 sensing target whose direction-independent RCS is reassigned at every 2--4~ms snapshot to the monostatic value of~\eqref{eq:rcs}; within a snapshot, motion enters through the target velocity. On an arm swing, Sionna and direct evaluation of~\eqref{eq:channel} have correlation 0.99998 in the time-varying component and power ratio 1.0001. This checks implementation consistency, not measured-channel validity. Monte Carlo runs therefore use the direct evaluation (about 0.1~s per episode versus one minute in Sionna), while Sionna is kept for these checks and for the multipath room; implementations and tests are public.\footnotemark[1] Fig.~\ref{fig:models} shows the single-point target's slow torso component and additional articulated limb components up to 4~m/s, using the same Sionna solver.

Preliminary experiments found near-chance event discrimination from 16 consecutive occasions over 8~ms; spreading the same 16 occasions over 32--64~ms raised window-level area under the receiver operating characteristic curve (AUC) to 0.86--0.88, favouring longer looks for these scenarios.

The first second of each episode is a warm-up without events or evaluation. Separate sets are used for training the LLR models (200 episodes), validation (200) and the main test (300, including 148 event-containing episodes). Validation selects parameters for $b\in\{2.5\%,5\%,10\%\}$ from $(L,d)\in\{(8,4),\allowbreak(16,4),\allowbreak(16,8),\allowbreak(32,4)\}$, uniform windows $N_w\in\{8,16,32\}$ and truncation levels $h\in\{2,5,10\}$. The recovery step $\mu$ is set so that adaptive monitoring cost matches the tuned periodic burst's realised cost within 5\%. Threshold $A$ is the smallest meeting the validation target of at most one false alarm per minute.

Parameters and acceptance criteria were frozen before test generation. Pre-registered adaptive-versus-periodic criteria required a $\Pfail$ reduction of at least 0.10 for events lasting up to 0.6~s, a positive lower 95\% confidence bound on that reduction, test false-alarm rates $\le1.25$/min and monitoring-cost mismatch $\le10\%$. The test allows 1.25/min to absorb sampling noise around the 1/min validation target. The main test was evaluated once.

\subsection{Pre-Registered Comparison}

\begin{table*}[!t]
\centering

\definecolor{isacNavy}{RGB}{36,59,83}
\definecolor{isacHeaderTint}{RGB}{233,239,246}
\definecolor{isacAdaptiveGray}{RGB}{244,246,249}
\definecolor{isacBestGreen}{RGB}{218,237,224}
\definecolor{isacMuted}{RGB}{133,141,151}
\definecolor{isacRule}{RGB}{169,180,192}

\newcommand{\isacBest}[1]{%
  \cellcolor{isacBestGreen}\textbf{#1}}
\newcommand{\isacNA}{%
  \textcolor{isacMuted}{---}}
\newcommand{\isacMetric}[2]{%
  \shortstack{\rule{0pt}{9pt}#1\\[-0.5pt]
  \rule[-2pt]{0pt}{7pt}$#2$}}
\newcommand{\isacHead}[2]{%
  \textcolor{white}{%
    \shortstack{\rule{0pt}{11pt}\textbf{#1}\\[2pt]
    \rule[-3pt]{0pt}{9pt}{\scriptsize #2}}}}

\caption{Sensing-schedule performance on three separate evaluation sets.}
\label{tab:results}

\footnotesize
\setlength{\tabcolsep}{2.4pt}
\setlength{\extrarowheight}{0pt}
\renewcommand{\arraystretch}{1.05}
\arrayrulecolor{isacRule}

\begin{tabular*}{\textwidth}{@{}ll
  @{\extracolsep{\fill}\hspace{6pt}}
  c@{\extracolsep{0pt}\hspace{2\tabcolsep}}*{4}{c}
  @{\extracolsep{\fill}\hspace{6pt}}
  c@{\extracolsep{0pt}\hspace{2\tabcolsep}}*{4}{c}
  @{\extracolsep{\fill}\hspace{6pt}}
  c@{\extracolsep{0pt}\hspace{2\tabcolsep}}*{3}{c}@{}}

\rowcolor{isacNavy}
\multicolumn{2}{@{}c@{\hspace{6pt}}}{%
  \raisebox{6pt}{\textcolor{white}{\textbf{Sensing schedules}}}}
 & \multicolumn{5}{@{}c@{\hspace{6pt}}}{%
     \isacHead{(a) Main free-space test}
              {300 episodes, 148 events}}
 & \multicolumn{5}{@{}c@{\hspace{6pt}}}{%
     \isacHead{(b) Two-rate trigger comparison}
              {300 fresh episodes, 140 events}}
 & \multicolumn{4}{@{}c@{}}{%
     \isacHead{(c) Multipath transfer}
              {200 episodes, 103 events}} \\

\rowcolor{isacHeaderTint}
\textbf{Budget} & \textbf{Schedule}
 & \isacMetric{$P_{\mathrm{fail}}$}{\downarrow}
 & \isacMetric{Timely(\#)}{\uparrow}
 & \isacMetric{FA/min}{\downarrow}
 & \isacMetric{Occ./s}{\downarrow}
 & \isacMetric{Delay (ms)}{\downarrow}
 & \isacMetric{$P_{\mathrm{fail}}$}{\downarrow}
 & \isacMetric{Timely(\#)}{\uparrow}
 & \isacMetric{FA/min}{\downarrow}
 & \isacMetric{Occ./s}{\downarrow}
 & \isacMetric{Delay (ms)}{\downarrow}
 & \isacMetric{$P_{\mathrm{fail}}$}{\downarrow}
 & \isacMetric{Timely(\#)}{\uparrow}
 & \isacMetric{FA/min}{\downarrow}
 & \isacMetric{Occ./s}{\downarrow} \\
\midrule[0.5pt]

\multirow{4}{*}{\textbf{2.5\%}} & Uniform
 & 0.99 & 2 & 1.26 & 32.8 & 430
 & \isacNA & \isacNA & \isacNA & \isacNA & \isacNA
 & \isacNA & \isacNA & \isacNA & \isacNA \\

 & Periodic burst
 & 0.97 & 5 & 0.80 & 32.9 & 471
 & 0.95 & 7 & 0.80 & 33.1 & 484
 & 0.94 & 6 & 3.37 & 31.6 \\

 & Two-rate trigger
 & \isacNA & \isacNA & \isacNA & \isacNA & \isacNA
 & 0.87 & 18 & 1.08 & 33.7 & 276
 & \isacBest{0.78} & 23 & 2.42 & 34.8 \\

\rowcolor{isacAdaptiveGray}
 & \textbf{Adaptive}
 & \isacBest{0.86} & 21 & 0.17 & 33.7 & 290
 & \isacBest{0.86} & 20 & 0.17 & 32.1 & 310
 & \isacBest{0.78} & 23 & 0.61 & 40.6 \\

\cmidrule[0.35pt](lr){1-2}\cmidrule[0.35pt](lr){3-7}
\cmidrule[0.35pt](lr){8-12}\cmidrule[0.35pt](lr){13-16}

\multirow{4}{*}{\textbf{5\%}} & Uniform
 & 0.99 & 1 & 1.14 & 64.4 & 405
 & \isacNA & \isacNA & \isacNA & \isacNA & \isacNA
 & \isacNA & \isacNA & \isacNA & \isacNA \\

 & Periodic burst
 & 0.84 & 23 & 0.29 & 66.8 & 218
 & 0.92 & 11 & 0.34 & 67.4 & 259
 & 0.82 & 19 & 1.04 & 66.0 \\

 & Two-rate trigger
 & \isacNA & \isacNA & \isacNA & \isacNA & \isacNA
 & 0.63 & 52 & 0.51 & 67.2 & 177
 & \isacBest{0.50} & 52 & 0.78 & 68.1 \\

\rowcolor{isacAdaptiveGray}
 & \textbf{Adaptive}
 & \isacBest{0.56} & 65 & 0.40 & 69.1 & 163
 & \isacBest{0.54} & 64 & 0.51 & 69.5 & 161
 & 0.51 & 50 & 1.12 & 71.8 \\

\cmidrule[0.35pt](lr){1-2}\cmidrule[0.35pt](lr){3-7}
\cmidrule[0.35pt](lr){8-12}\cmidrule[0.35pt](lr){13-16}

\multirow{3}{*}{\textbf{10\%}} & Uniform
 & 0.53 & 70 & 1.31 & 125.0 & 179
 & \isacNA & \isacNA & \isacNA & \isacNA & \isacNA
 & \isacNA & \isacNA & \isacNA & \isacNA \\

 & Periodic burst
 & 0.53 & 69 & 0.40 & 127.1 & 177
 & \isacNA & \isacNA & \isacNA & \isacNA & \isacNA
 & 0.50 & 51 & 0.78 & 126.4 \\

\rowcolor{isacAdaptiveGray}
 & \textbf{Adaptive}
 & \isacBest{0.29} & 105 & 0.57 & 127.3 & 122
 & \isacNA & \isacNA & \isacNA & \isacNA & \isacNA
 & \isacBest{0.26} & 76 & 1.21 & 125.6 \\

\bottomrule[0.6pt]
\end{tabular*}
\arrayrulecolor{black} 

\par\vspace{2pt}
\begin{minipage}{\textwidth}
\footnotesize
\textit{Notes.}
$P_{\mathrm{fail}}$: fraction of missed events.
Timely: number of events detected within 200~ms of onset
and before motion ends.
FA/min: false alarms per minute of event-free time.
Occ./s: monitoring occasions per second, excluding service sensing.
Delay: median onset-to-alarm time for alarms during the event
or within 0.5~s after it ends.
Budgets are nominal requested-slot fractions.
Panel~(b) reports only 2.5\% and 5\%; Delay is omitted in~(c) for space.
In~(c), schedules are used without retuning; monitoring costs
and false-alarm rates are not re-matched.
Grey rows mark Adaptive; green cells mark the lowest
$P_{\mathrm{fail}}$ per budget and panel, including ties.
Dashes mean not evaluated or not reported.
\end{minipage}
\end{table*}

In Table~\ref{tab:results}(a), the adaptive schedule (labelled Adaptive) detects more events within the deadline and has lower median onset-to-alarm delay at every budget. Timely detections increase by factors 4.2, 2.8 and 1.5 over periodic bursts at 2.5\%, 5\% and 10\%. Pre-registered criteria are met at both budgets they cover (2.5\% and 5\%). For 93 events lasting $\le0.6$~s, paired $\Pfail$ reductions are 0.11 (95\% bootstrap confidence interval $[0.02,0.19]$) and 0.25 ($[0.16,0.34]$), respectively.

Adaptive and periodic-burst monitoring costs differ by $<4\%$, with false-alarm rates $<1$/min. On 200 additional event-free episodes, Adaptive produces 0.08 and 0.30/min at 2.5\% and 5\%.

\begin{figure}[t]
\centering
\includegraphics[width=\columnwidth]{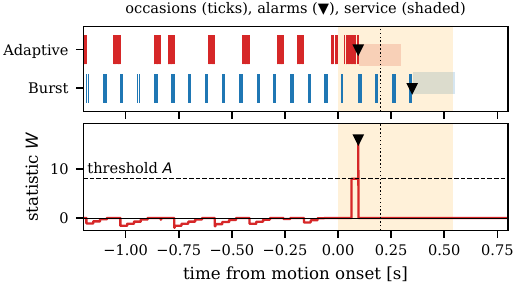}
\caption{One validation episode at 5\% budget (arm swing, shaded). Top: sensing occasions of each schedule. Bottom: the adaptive statistic $W$. The dotted line marks the 200~ms deadline.}
\label{fig:trace}
\end{figure}

In the validation episode of Fig.~\ref{fig:trace}, negative pre-onset evidence skips looks; positive evidence sustains consecutive looks and triggers an alarm 95~ms after onset. At comparable average monitoring cost, periodic bursts every 80~ms alarm 349~ms after onset, beyond the deadline.

\subsection{Ablation of Timing and Doppler Features}

At 5\%, the tuned periodic burst uses $8{\times}4$ looks and Adaptive $16{\times}4$, so part of the gain could come from the look rather than the timing. Table~\ref{tab:ablation} tests look design and Doppler through identical-look and energy-only comparisons. Identical $16{\times}4$ looks slightly increase periodic-burst failure, while Adaptive's advantage persists. Fresh sensing-rate sweep episodes (Section~\ref{sec:regime}), with identical looks and matched monitoring costs (ratios 0.98--1.02), give periodic failure rates of 98\% and 96\% at 2.5\% and 5\%, respectively, versus 84\% and 59\% for Adaptive.

The gain does not require Doppler: at 5\%, energy-only Adaptive gives $\Pfail=0.63$ versus 0.92 for tuned periodic bursts with both features. Doppler further reduces adaptive failure by 7--9 percentage points across budgets, with little periodic-burst change. A plausible reason is that consecutive looks capture the onset. These post-test analyses use one validation set without uncertainty estimates; the Doppler benefit is exploratory and the spectral-band rule heuristic.

\begin{table}[t]
\centering

\definecolor{isacAbNavy}{RGB}{36,59,83}
\definecolor{isacAbTint}{RGB}{233,239,246}
\definecolor{isacAbGray}{RGB}{244,246,249}
\definecolor{isacAbGreen}{RGB}{218,237,224}
\definecolor{isacAbRule}{RGB}{169,180,192}

\caption{Exploratory ablation of look design and features.}
\label{tab:ablation}

\footnotesize
\setlength{\tabcolsep}{3pt}
\newlength{\isacAbWidth}
\setlength{\isacAbWidth}{%
  \dimexpr\columnwidth-8\tabcolsep\relax}
\setlength{\extrarowheight}{0pt}
\renewcommand{\arraystretch}{1.05}
\arrayrulecolor{isacAbRule}

\begin{tabular}{@{}
  >{\raggedright\arraybackslash}p{0.27\isacAbWidth}
  >{\centering\arraybackslash}p{0.15\isacAbWidth}
  >{\centering\arraybackslash}p{0.34\isacAbWidth}
  >{\centering\arraybackslash}p{0.12\isacAbWidth}
  >{\centering\arraybackslash}p{0.12\isacAbWidth}@{}}

\rowcolor{isacAbNavy}
\multicolumn{3}{@{}c}{%
  \rule[-3pt]{0pt}{14pt}%
  \textcolor{white}{\textbf{Configuration}}}
 & \multicolumn{2}{c@{}}{%
   \textcolor{white}{$P_{\mathrm{fail}}\;\downarrow$}} \\

\rowcolor{isacAbTint}
\textbf{Schedule} & \textbf{Look} & \textbf{Features}
 & \textbf{2.5\%} & \textbf{5\%} \\
\midrule[0.5pt]

Periodic burst & Tuned & Energy + Doppler
 & 0.95 & 0.92 \\

Periodic burst & $16{\times}4$ & Energy + Doppler
 & 0.99 & 0.93 \\

Periodic burst & Tuned & Energy
 & 0.97 & 0.92 \\

\cmidrule[0.35pt](lr){1-3}
\cmidrule[0.35pt](lr){4-5}

\rowcolor{isacAbGray}
\textbf{Adaptive} & $16{\times}4$ & Energy + Doppler
 & \cellcolor{isacAbGreen}\textbf{0.81}
 & \cellcolor{isacAbGreen}\textbf{0.56} \\

\rowcolor{isacAbGray}
\textbf{Adaptive} & $16{\times}4$ & Energy
 & 0.90 & 0.63 \\

\bottomrule[0.6pt]
\end{tabular}
\arrayrulecolor{black}

\par\vspace{2pt}
\begin{minipage}{\columnwidth}
\footnotesize
\textit{Notes.} Look is $L{\times}d$: $L$ requested occasions,
$d$ slots apart. Tuned uses the validation-selected design.
Post-test analysis on validation data; alarm thresholds
and monitoring costs are recalibrated for each configuration.
Budgets are nominal requested-slot fractions.
Grey rows mark Adaptive; green cells mark the lowest
$P_{\mathrm{fail}}$ per budget.
\end{minipage}
\end{table}
\subsection{Comparison with a Two-Rate Trigger}

The two-rate trigger was added after the main test and separately pre-registered on 300 fresh episodes (140 events), alongside frozen periodic-burst and adaptive schedules. Validation selected its look design, $\tau\in\{0,2,4\}$ and $N_{\mathrm{hold}}\in\{2,4,8\}$; $P_{\mathrm{slow}}$ matched periodic-burst monitoring cost.

The trigger improves on periodic bursts (Table~\ref{tab:results}(b)). Paired $\Pfail(\text{trigger})-\Pfail(\text{adaptive})$ differences are $+0.01$ (95\% confidence interval $[-0.06,0.09]$) at 2.5\% and $+0.09$ ($[0.00,0.18]$) at 5\%. Rates are close at 2.5\%; at 5\%, Adaptive fails less often on average, but the confidence interval reaches zero. The two schedules thus perform comparably at the sample sizes used. At 2.5\%, however, Adaptive reaches the same failure rate with far fewer false alarms (0.17 versus 1.08/min; 0.61 versus 2.42/min in the room), while at 5\% they are similar.

Unlike the trigger, Adaptive needs no separate switching threshold or hold length (12 versus 36 grid configurations per budget), although calibration effort was not measured.

\subsection{Dependence on Sensing Rate}\label{sec:regime}

\begin{figure}[t]
\centering
\includegraphics[width=\columnwidth]{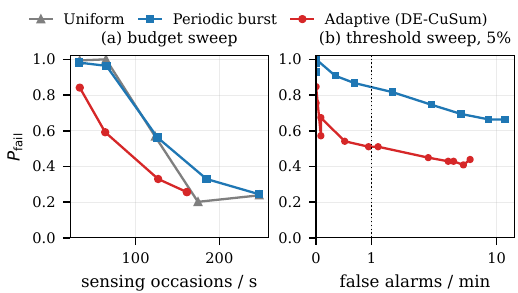}
\caption{(a) $\Pfail$ versus realised monitoring rate on fresh episodes, using look $16{\times}4$ and a validation false-alarm target of $\le1$/min. (b) Validation failure probability versus false-alarm rate under an alarm-threshold sweep at the 5\% nominal budget, with the LLR clip widened to $[-20,20]$ so that the sweep is continuous.}
\label{fig:results}
\end{figure}

The sensing-rate sweep fixes look $16{\times}4$, calibrates nominal budgets 2.5--20\% on validation data and uses 300 fresh reserved episodes. Adaptive has lower failure up to about 130 occasions/s; above roughly 170, Uniform achieves comparable or lower failure (Fig.~\ref{fig:results}a).

Aliasing of Uniform may contribute, since limb speeds up to 4.5~m/s imply Doppler shifts up to $\pm105$~Hz at 3.5~GHz, but detection does not require alias-free reconstruction. Moreover, Adaptive skips at least one epoch after negative evidence, capping its monitoring near 160 occasions/s, so its two highest-budget points cost less than the baselines and the crossover is not a cost-matched comparison.

Sweeping alarm threshold $A$ at the 5\% budget (Fig.~\ref{fig:results}b), Adaptive has lower failure than periodic bursts across the plotted false-alarm range, from zero to more than ten per minute.

\subsection{Transfer to a Multipath Room}

To test whether the gain survives multipath, the schedules were moved without retuning into a closed $8{\times}10{\times}3$~m concrete room simulated in Sionna~RT on an NVIDIA H100 GPU. Target echoes now include up to two specular wall reflections, and the static background is about 39~dB stronger than the human echo. Monitoring costs and false-alarm rates were not re-matched, and the 200 room episodes were evaluated under a separate pre-registration (Table~\ref{tab:results}(c)).

Adaptive still outperforms periodic bursts at all budgets: for events lasting $\le0.6$~s, the paired $\Pfail$ reductions are 0.14 ($[0.02,0.25]$), 0.31 ($[0.17,0.44]$) and 0.22 ($[0.10,0.34]$) at 2.5\%, 5\% and 10\%. The trigger and Adaptive again perform similarly, with failure differences of $0.00$ ($[-0.10,0.10]$) and $-0.02$ ($[-0.12,0.07]$) at 2.5\% and 5\%. Uniform, not shown in the table, misses 52--99\% of events with 2.1--2.9 false alarms/min.

The stronger clutter shifts the operating points. Adaptive false alarms exceed 1/min at 5\% and 10\%, and its monitoring cost is 1.29 and 1.09 times that of periodic bursts at 2.5\% and 5\%, and about equal at 10\%. With thresholds recalibrated on 100 separate room episodes, Adaptive still fails 0.20--0.25 less often than periodic bursts, though at 5\% costs stay unmatched (ratio 1.13) and periodic false alarms reach 3.5/min.

\section{Discussion and Limitations}

The results point to a clear operating regime. Evidence-driven timing earns its keep when the monitoring budget is tight; once the budget is generous, sensing at a steady rate is simpler and performs just as well. Two side effects deserve mention. First, equal monitoring cost does not imply equal total cost: because the adaptive schedule catches more events, it also starts more post-alarm service sensing, which raises its total cost by 20\% over periodic bursts at the 5\% budget. Second, the room shows what happens when thresholds tuned in free space meet much stronger clutter: false alarms rise and costs no longer match. This stems from moving the thresholds without recalibration, not from the scheduling rule itself.

These findings rest on simulation, and several simplifications should be kept in mind. The base station has a single antenna, the body is built from ellipsoids, busy slots occur independently, and the scheduler runs inside the DU without higher-level radio access network (RAN) delays. In the room, reflected paths reuse the RCS seen along the direct path. For irregular samples the Doppler band is set heuristically, so the added value of the Doppler feature remains exploratory. The lowest budget is also hard in absolute terms: at 2.5\%, even the best schedule misses most events, as expected with one antenna, a 5~dB torso SNR and a 200~ms deadline, and an antenna array should lower all curves. Finally, the body model has been checked against a second implementation, but not yet against measured channels.

\section{Conclusion}

Sparse ISAC sensing can miss brief limb motion, and the standard single-point human model cannot even represent it. With an articulated body model and DE-CuSum adapted to OFDM looks, evidence-driven timing detects up to four times as many brief movements in time as periodic sensing at matched monitoring cost; a tuned trigger performs comparably, and the gain over periodic sensing survives transfer to a multipath room. Future work will validate the approach on an OpenAirInterface 5G testbed and implement it as a DU application in an Open RAN stack.

\section*{Acknowledgment}
\noindent
This material is based upon work supported in part by NSF under Awards CNS-2120442 and IIS-2325863, and NTIA under Award No. 51-60-IF007. Any opinions, findings, and conclusions or recommendations expressed in this publication are those of the author(s) and do not necessarily reflect the views of the NSF and NTIA. AI assisted with code and text; the authors verified all content.

\bibliographystyle{IEEEtran}
\bibliography{refs}

\end{document}

%% file: figures/fig_system.tex
\begin{tikzpicture}[
  font=\scriptsize, >=Stealth,
  box/.style={draw, rounded corners=1.5pt, fill=#1, minimum height=5.2mm, inner xsep=2pt, align=center},
  box/.default=gray!8,
  occ/.style={fill=red!75!black, draw=none},
  busy/.style={fill=gray!45, draw=none},
  free/.style={fill=white, draw=gray!50, line width=0.2pt},
]
\node[font=\scriptsize\bfseries, anchor=west] at (-0.62,2.55) {(a)};
\draw[thick] (0.35,0.55) -- (0.55,1.65) -- (0.75,0.55);
\draw (0.42,0.95) -- (0.68,0.95); \draw (0.47,1.25) -- (0.63,1.25);
\fill[blue!60!black] (0.45,1.65) rectangle (0.65,2.05);
\node[align=center, anchor=north] at (0.55,0.5) {gNB\\[-1pt](full duplex)};
\draw[->, thick, blue!60!black, decorate, decoration={snake, amplitude=0.6mm, segment length=2.2mm, post length=1.2mm}]
  (0.85,1.95) -- (4.85,1.55) node[pos=0.48, above=1.5pt, black] {OFDM sensing occasion};
\draw[->, thick, red!70!black, dashed] (4.85,1.2) -- (0.85,1.6) node[pos=0.5, below=1pt, black] {echo $y_k[n]$ (Sionna cross-checked)};
\foreach \x/\y in {2.35/0.55, 3.25/0.75, 6.85/0.6} \fill[gray!55] (\x,\y) rectangle ++(0.22,0.22);
\node[gray!70!black] at (2.9,0.35) {static clutter, drift};
\begin{scope}[shift={(5.55,0.25)}, line width=0.9pt, line cap=round]
  \draw (0,1.62) circle (0.14);                
  \draw (0,1.48) -- (0,0.85);                  
  \draw (0,0.85) -- (-0.17,0.3); \draw (0,0.85) -- (0.17,0.3);   
  \draw (0,1.35) -- (-0.22,0.98);              
  \draw[red!70!black] (0,1.35) -- (0.42,1.72); 
  \draw[->, red!70!black, line width=0.5pt] (0.62,1.25) arc[start angle=-40, end angle=55, radius=0.42];
\end{scope}
\node[align=center, anchor=west] at (6.3,1.45) {limb moves,\\[-1pt]torso still};
\node[align=center, anchor=north] at (5.55,0.55) {12-segment\\[-1pt]body};

\begin{scope}[yshift=-0.85cm]
\node[font=\scriptsize\bfseries, anchor=west] at (-0.62,0.25) {(b)};
\def\N{32}
\foreach \i in {0,...,31} {
  \pgfmathtruncatemacro{\b}{(mod(\i*7+3,10)<3)}  
  \ifnum\b=1 \fill[busy] (0.25+\i*0.235,-0.75) rectangle ++(0.205,0.32);
  \else \draw[free] (0.25+\i*0.235,-0.75) rectangle ++(0.205,0.32); \fi
}
\foreach \j in {2,4,6,8} {
  \pgfmathtruncatemacro{\bb}{(mod(\j*7+3,10)<3)}
  \ifnum\bb=0 \fill[occ] (0.25+\j*0.235,-0.75) rectangle ++(0.205,0.32);
  \else \draw[red!75!black, line width=0.5pt] (0.25+\j*0.235,-0.75) -- ++(0.205,0.32) (0.25+\j*0.235+0.205,-0.75) -- ++(-0.205,0.32); \fi
}
\draw[decorate, decoration={brace, amplitude=1.5mm}] (0.25+2*0.235,-0.35) -- (0.25+9*0.235-0.03,-0.35)
  node[midway, above=1.6mm, align=center] {look: $L$ occasions, spacing $d$\\[-1pt]span $LdT_s$ $\to$ resolution};
\draw[decorate, decoration={brace, amplitude=1.5mm}] (0.25+11*0.235,-0.35) -- (0.25+22*0.235-0.03,-0.35)
  node[midway, above=1.6mm] {skip while $W<0$};
\foreach \j in {23,25,27,29} {
  \pgfmathtruncatemacro{\bb}{(mod(\j*7+3,10)<3)}
  \ifnum\bb=0 \fill[occ] (0.25+\j*0.235,-0.75) rectangle ++(0.205,0.32);
  \else \draw[red!75!black, line width=0.5pt] (0.25+\j*0.235,-0.75) -- ++(0.205,0.32) (0.25+\j*0.235+0.205,-0.75) -- ++(-0.205,0.32); \fi
}
\draw[->] (0.2,-0.95) -- (7.9,-0.95) node[right=-1pt] {slot};
\fill[occ] (0.3,-1.27) rectangle ++(0.16,0.16); \node[anchor=west] at (0.48,-1.19) {sensing occasion};
\fill[busy] (2.35,-1.27) rectangle ++(0.16,0.16); \node[anchor=west] at (2.53,-1.19) {data slot (30\%)};
\draw[red!75!black, line width=0.5pt] (4.45,-1.27) -- ++(0.16,0.16) (4.61,-1.27) -- ++(-0.16,0.16);
\node[anchor=west] at (4.63,-1.19) {request rejected};

\end{scope}
\begin{scope}[yshift=-0.95cm]
\node[font=\scriptsize\bfseries, anchor=west] at (-0.62,-1.7) {(c)};
\node[box] (fft) at (0.95,-2.25) {range\\[-1pt]FFT};
\node[box] (bg)  at (2.2,-2.25) {background\\[-1pt]removal};
\node[box] (ft)  at (3.6,-2.25) {energy +\\[-1pt]Doppler};
\node[box] (llr) at (4.75,-2.25) {LLR\\[-1pt]$\ell$};
\node[box=red!8] (w) at (6.35,-2.25) {DE-CuSum $W$:\\[-1pt]look / skip / alarm};
\draw[->] (fft) -- (bg); \draw[->] (bg) -- (ft); \draw[->] (ft) -- (llr); \draw[->] (llr) -- (w);
\node[box=blue!6] (sch) at (3.6,-3.25) {scheduler: accept / reject};
\draw[->] (w.south) |- node[pos=0.62, below=0.5pt] {request, $+1$ slot} (sch.east);
\draw[->] (sch.west) -| node[pos=0.42, below=0.5pt] {delivery, $+2$ slots} (fft.south);
\end{scope}
\end{tikzpicture}